\documentclass[intlimits,twoside,a4paper]{article}

\usepackage{amsmath,amssymb}
\usepackage{graphicx}
\usepackage{subfigure}

\usepackage[T2A]{fontenc}
\usepackage[cp1251]{inputenc}

\usepackage[eqsecnum]{cmpj2}

\issue{2014}{17}{1}{13002}
\doinumber{10.5488/CMP.17.13002}

\title[A simple model potential for hollow nanospheres]%
{A simple model potential for hollow nanospheres}
\author[K. K\"{o}ksal \textsl{et al.}]{K.~K\"{o}ksal\refaddr{Bitlis},
        M.~\"{O}ncan\refaddr{Bitlis, Antep}, Bulent~G\"{o}n\"{u}l\refaddr{Antep}, Besire~G\"{o}n\"{u}l\refaddr{Antep}}

\authorcopyright{K. K\"{o}ksal, M.~\"{O}ncan, Bulent~G\"{o}n\"{u}l, Besire~G\"{o}n\"{u}l, 2014}

\date{Received October 7, 2013, in final form November 27, 2013}
        \addresses{
\addr{Bitlis} Physics Department, Bitlis Eren University, 13000,
Bitlis, Turkey
\addr{Antep} Department of Engineering Physics, Gaziantep University, 27100,
Gaziantep, Turkey
}

\begin{document}

\maketitle

\begin{abstract}
A new model potential is introduced to describe the hollow
nanospheres such as fullerene and molecular structures and to obtain
their electronic properties. A closed analytical solution of the
corresponding treatment is given within the framework of supersymmetric
perturbation theory.
\keywords electronic structure, model potential, analytical solution
\pacs 03.65.Ge, 31.15.B-, 31.15.xp
\end{abstract}

\section{Introduction}

Nanostructures have received great interest nowadays
because of their importance in solid state technology and for
medical purposes \cite{Koray2,Lauren,Li,Ahn,salata}.
Specifically, quantum dots have been studied to strongly confine and
control the electrons in a few nanometers of the volume \cite{akgul,
sahin}. Nevertheless, semiconductor quantum dots can be produced using
high technological growth devices while the production of metallic
quantum dots or clusters also requires advanced technology
\cite{harrison}. However, molecular nanospheres can be obtained
more easily (e.g., by chemical synthesis) compared with those of
metal and semiconductor quantum dots~\cite{guldi2002fullerenes}.

Molecular nanoparticles as in the case of metal clusters \cite{Ma,
Podsiadlo} can be produced in the shape of hollow spheres. In these
structures, there is a limited radial motion of the electrons. As an
example of hollow nanospheres, we can consider the fullerene
molecule which is stable and consists of $60$ carbon atoms.
The derivatives of fullerene are C$_{70}$,C$_{84}$, C$_{540}$
\cite{pavlyukh1, pavlyukh2}. Furthermore, B$_{80}$ is a spherical
molecule in which one uses the boron atoms instead of carbon
\cite{zhao}.

For complex molecular systems which include more than ten atoms,
the best way of electronic structure analysis is to use some
ab-initio \cite{scuseria, gonzalez} or semi-empirical techniques
\cite{Halac, chen}. Although these are known as realistic
calculations, they have time-consuming procedures and give only
numerical results. However, since the molecules like fullerene have
spherical symmetry, if the motion of an electron can be described in
only one dimension (which is radial direction), the whole motion will be
described due to the spherical symmetry. Appropriate model
potentials can describe the motion of the electrons in radial
direction \cite{Dolmatov, George, Grzhimailo, Nascimento, Lin}. The
angular motion of an electron is described by well known spherical
harmonics. In other words, the related spherical symmetric
one-dimensional radial Schr\"{o}dinger equation
analytically yields the spectra of interest, from which one can readily
observe the behaviours and the physics behind the system considered,
unlike its corresponding numerical results.

Within the context, we remind that the attractive Gaussian potential
\begin{equation}
  V(\lambda;r)=-\gamma \re^{-\lambda r^2}
\label{gaussian}
\end{equation}%
is very important in modelling the nucleon-nucleon scattering
\cite{Buck1977246} and quantum dots with single or more electrons
\cite{Gomez, Adamowski, Diego}, impurity \cite{Wenfang2010239,
Lu20114129} and excitons \cite{Hours}. In equation~\eqref{gaussian},
$\gamma$ shows the height of the radial central potential, $\lambda$
is related to the dot radius ($R$) in such a way that
$\lambda=1/R^2$.

We have recently obtained a closed form of the eigenvalues of this
potential \cite{koksal} and shown that the obtained results are very accurate,
particularly for low principle quantum numbers. For completeness,
in this article we aim to suggest that such interaction profile,
apparently with a plausible modification, should be also suitable for the
investigation of hollow nanospheres.

The paper is organized as follows. The next section briefly discusses
the model potential. Theory section includes the calculation procedure
and the discussion on the results obtained. Some concluding remarks and
outlook are given in the final section.

\section{Model potential}

\begin{figure}[!ht]
  \centerline{\includegraphics[width=8cm]{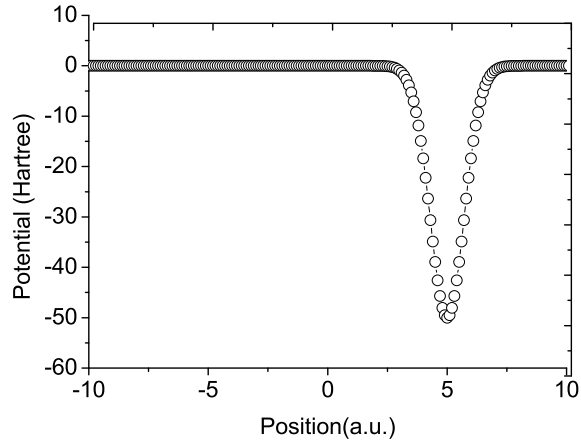}}
  \caption{Attractive gaussian potential which has a formula given by equation~\eqref{gaussiannas}.
  As can be understood from the shape, the potential is  not symmetric and unsuitable for analytical solution. Here, the $a.u.$
  indicates atomic unit and should be considered as Bohr radius.}\label{Fig:oldpot}
\end{figure}

The electronic properties of a spherical symmetric molecular cluster
can be obtained by modelling this structure. So far, many of the
studies have been performed in the framework of numerical calculations.
In previous studies regarding fullerene and endohedral structures,
the spherical potential which models the structures is presented in
different forms \cite{Dolmatov, George, Grzhimailo, Nascimento,
Lin}. The square potential having the following form
\begin{eqnarray}
V(r)&=&\left\{
\begin{array}{ll}
  - V_0 &\quad  \text{if} \quad r_0\leqslant  r \leqslant  r_i+d,\\
  0, &\quad  \text{if} \quad  r<r_0\,,
\end{array}
\right.
\\\nonumber
  &&\text{$d$: the thickness of the wall},\\\nonumber
  &&\text{$r_0$: inner radius}
\label{sqwpot}
\end{eqnarray}%
has been introduced by \cite{George, Grzhimailo} to describe the
fullerene. To describe an atom inside C$_{60}$, Dolmatov et. al.
has introduced the Woods-Saxon potential \cite{Dolmatov} as
follows:
\begin{equation}
  V(r)=\left[\frac{2V_0}{1+\exp{\left(\frac{r_0-r}{\nu}\right)}}\right]_{r\leqslant
  r_0+\Delta/2}+\left[\frac{2V_0}{1+\exp{\left(\frac{r-r_0-r}{\nu}\right)}}\right]_{r>
  r_0+\Delta/2}\,,
\label{woodssaxon}
\end{equation}%
where $\nu$ is a parameter related to the additional atom, $r_0$,
$V_0$ and $\Delta$ are potential parameters. Lin et. al. have used a
power-exponential potential which has the following form \cite{Lin}
\begin{equation}
  V(\lambda;r)=-\gamma \exp\left[-\lambda(r-r_0)^p/w^p\right],
\label{pow-exp}
\end{equation}%
where $r_0$ and $w$ indicate the radius and and thickness of the
hollow cage. Nascimento et. al. have introduced the gaussian-type
potential for modelling the fullerene-type structures as follows
\cite{Nascimento}
\begin{equation}
  V(\lambda;r)=-\gamma \exp\left[-\lambda(r-r_0)^2\right].
\label{gaussiannas}
\end{equation}%
Using the mentioned potentials, it is possible to obtain
electronic structures of hollow spheres. As an example, the last
potential has the shape as seen in figure~\ref{Fig:oldpot}. Since the
parameter $r_0$ breaks the symmetry and makes the potential
discontinuous, the solution of the potential can be performed only
by using the numerical techniques.
An analytical solution for a potential, $r_0$, should be zero.
Therefore, a spherical hollow nanostructure may be described by the
combination of an attractive Gaussian potential with an additional
term as follows
\begin{equation}
  V(\lambda;r)=-\gamma \re^{-\lambda r^2}+\frac{\beta}{r^2}\,,
\label{gaussian2}
\end{equation}%
where the second term including $\beta$ (this should be positive) is
very similar to the barrier term due to the spherical symmetry which
is $\ell (\ell+1)/r^2$ \cite{aygun}. In the case of $\beta=0$, the
potential in equation~\eqref{gaussian2} can describe a solid nanoparticle
such as metal clusters, while for $\beta \neq 0$, a hollow nanosphere
can be described.

\begin{figure}[!ht]
  \centerline{\includegraphics[width=8cm]{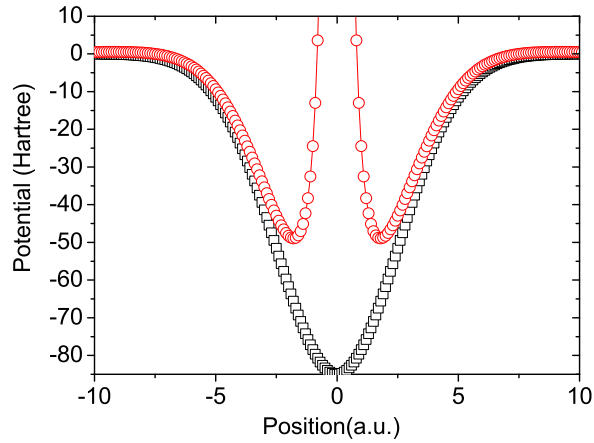}}
  \caption{(Color online) The model potential to describe a hollow nanosphere.
  The red circles indicate the exponential part of equation~\eqref{gaussian2} and the black squares show the
  second part in  equation~\eqref{gaussian2}. }\label{Fig:newpot}
\end{figure}
Figure~\ref{Fig:newpot} shows the potential introduced in equation~\eqref{gaussian2}. Due to the additional term
(${\beta}/{r^2}$), the electrons in the system will be pushed to
make a motion in the regions far from the center of the sphere. The
increase in the $\beta$ will cause a decrease of the motion area
of the electrons. Therefore, this potential can describe a group of
spherical symmetric molecular structures and different types of
hollow nanospheres.

\section{Theory}

As seen from equation~\eqref{gaussian2}, the potential is constructed
with two parts. $\beta/r^2$ can be introduced into the barrier term
$-\ell(\ell+1)/r^2$ following the previous study by Aygun et al.
\cite{aygun}. In this case, we can describe a new parameter
$\ell'(\ell'+1)=-\ell(\ell+1)+\beta$ which results in:
\begin{equation}
  \ell'=\frac{1}{2}\left(-1 + \sqrt{1+4\ell+4\ell^2+4\beta}\right)\,.
\label{barrierterm}
\end{equation}
The attractive Gaussian potential part in equation~\eqref{gaussian2} can be
expanded as follows
\begin{equation}
  V(\lambda; r)=-\gamma+ \gamma \lambda r^2-\frac{\gamma\lambda^2}{2} r^4+\dots ,
\label{expgaussian}
\end{equation}
where $\gamma \lambda r^2$ indicates the well known harmonic
oscillator potential where $\gamma \lambda=m\omega^2/2$. The role of
the terms other than $-\gamma+ \gamma \lambda r^2$ is to modify the
potential and they may be referred to as modification potentials. The
normalized wavefunction of a harmonic oscillator is
\begin{equation}
    \chi_{n\ell'}(r)=\sqrt{\frac{2^{n+2\ell'+3+1/2}}{\sqrt{\pi}}\frac{n!}{(2n+2\ell'+1)!!}}\left(\frac{\lambda\gamma m}{2\hbar^2}\right)
    ^{\frac{2\ell'+3}{8}}\re^{-\sqrt{\frac{\lambda\gamma m}{2\hbar^2}} r^2}L_n^{\ell'+1/2}\left(\sqrt{2\lambda\gamma m}r^2/\hbar\right) r^{\ell'+1}\label{harmonic}\,,
\end{equation}%
where $m$ and $\hbar$ are electron mass and Planck's constant,
respectively, $L_n^{\ell'+1/2}\left(\!\sqrt{2\lambda\gamma m}r^2/\hbar\right)$ is
associated Laguerre polynomial.

Here, the corresponding eigenvalue and superpotential terms of the
harmonic oscillator term can be written from the literature
\cite{Dabrowska}
\begin{eqnarray}
    W_{n=0}(r)&=&-\frac{\hbar}{\sqrt{2m}}\frac{\chi'_{n\ell'}(r)}{\chi_{n\ell'}(r)}=-\frac{\hbar}{\sqrt{2m}}\frac{\ell'+1}{r}+\sqrt{\gamma\lambda}\, r,\nonumber\\
    E_{n,\ell'}&=&\hbar\sqrt{\frac{2\gamma\lambda}{m}}(2 n+\ell'+3/2).
    \label{parabolicenergy}
\end{eqnarray}
Perturbative wavefunctions, energies and superpotentials
corresponding to the modification potentials are as follows:
\begin{eqnarray}
\Delta V(r;\epsilon)&=&\sum_{k=1}^{\infty}\epsilon^k\Delta V^{\{k\}}(r),\nonumber\\
    \Delta W_{n\ell'}(r;\epsilon)&=&\sum_{k=1}^{\infty}\epsilon^k\Delta W_{n \ell'}^{\{k\}}(r),\nonumber\\
    \Delta E_{n\ell'}(\epsilon)&=&\sum_{k=1}^{\infty}\epsilon^k\Delta E_{n \ell'}^{\{k\}}\,,
 \label{perturbation}
  \end{eqnarray}
where $k$ indicates the perturbation order. If the unknown perturbed wavefunction is $R_{\mathrm{P}}(r)$, the Schr\"{o}dinger
equation can be written as follows:
\begin{equation}
-\frac{\hbar^2}{2m}\left[\frac{\chi''_{n\ell'}}{\chi_{n\ell'}}+\frac{R''_{\mathrm{P}}(r)}{R_{\mathrm{P}}(r)}+
2\frac{\chi'_{n\ell'}}{\chi_{n\ell'}}\frac{R'_{\mathrm{P}}(r)}{R_{\mathrm{P}}(r)}
    \right]=V_H+V_{\mathrm{P}}+E_{n\ell'}+\Delta E_{n\ell'}\,.
 \label{schrodinger}
\end{equation}%
Skipping the details of the calculation procedure of the theory to
our previous work \cite{koksal}, the total energy eigenvalues can be
written as follows:
\begin{equation}
    \Im_{n,\ell}=\left(2n+\frac{-1 + \sqrt{1+4\ell+4\ell^2+4\beta}}{2}+\frac{3}{2}\right)\sqrt{\frac{\gamma\lambda \hbar^2}{2m}}
        -\gamma \re^{-{\left[2n+\frac{1}{2}(-1 + \sqrt{1+4\ell+4\ell^2+4\beta})+3/2\right]} \sqrt{\frac{\lambda\hbar^2}{2\gamma m}} },
\end{equation}%
where $\Im_{n,\ell}=E_{n\ell'}+\Delta E_{n\ell'} \ \big(\ell'\rightarrow
\big[-1 + \sqrt{1+4\ell+4\ell^2+4\beta}\big]/2\big)$. Therefore, we obtain a
simple analytical form of the energy eigenvalues.

\begin{table}
  \centering
  \caption{The eigenvalues of the potential for $\ell=0$ and for different parameters in atomic units.
  The value of the $\gamma$ is 4 Hartree. One should note that $n$ is a radial quantum number.  }\label{tabl0}
  \vspace{2ex}
  {\small
\begin{tabular}{|l||l|l|l|l|l|l|l|l|l|}
  \hline
$n$  & &$\lambda=0.0125$& & &$\lambda=0.025$&&&$\lambda=0.0125$ &\\
\hline\hline
&$\beta=0$ & $\beta=2.5$&$\beta=5$&$\beta=0$& $\beta=2.5$& $\beta=5$&$\beta=0$& $\beta=2.5$&$\beta=5$\\
 \hline 0 &   --2.9404&
--1.79436 &   --1.20055  &  --3.37972 &   --2.69002 &--2.32524 &--3.55892
&   --3.06275  &  --2.79796\\ \hline 1 &  --1.94631& --1.30392& --0.83482
&   --2.78255 &   --2.3891& --2.09817 & --3.12967& --2.84443& --2.63231\\
\hline 2 &   --1.00933  &  --0.5972& --0.24876&
--2.20674& --1.94968& --1.73062 &   --2.71159 &   --2.52366  & --2.36291\\
\hline 3& --0.12217& & 0.4336 & --1.65066  &  --1.46551& --1.2973&
--2.30409 &--2.16762& --2.0433\\ \hline 4   & &  && --1.11281& --0.9707&
--0.8373& --1.90659& --1.80102  &  --1.70174 \\  \hline 5 && &
&--0.59181& --0.47792& --0.36901  & --1.51854& --1.43333& --1.35173
\\ \hline 6& & & & --0.08637 && &--1.13943& --1.06854& --0.99989\\ \hline
7 & & &&& & &--0.76877 &--0.70845& --0.6496\\ \hline 8 & &&&& & &
--0.40612& --0.35388& --0.30266\\ \hline  9 &   &  &  & & &  &
--0.05102 &  --0.00515 &  \\
  \hline
\end{tabular}}
\end{table}

\begin{table}
  \centering
  \caption{The eigenvalues of the potential for $\ell=1$ and for different parameters in atomic units.
  The value of the $\gamma$ is 4 Hartree. }\label{tabl1}
  \vspace{2ex}
{\small
\begin{tabular}{|l||l|l|l|l|l|l|l|l|l|}
  \hline
$n$  & &$\lambda=0.0125$& & &$\lambda=0.025$&&&$\lambda=0.0125$ &\\
\hline\hline
&$\beta=0$ & $\beta=2.5$&$\beta=5$&$\beta=0$& $\beta=2.5$& $\beta=5$&; $\beta=0$& $\beta=2.5$&$\beta=5$\\
 \hline
0  &  --1.00933 &  && --2.20674 &   --1.56434   & --1.22363&
--2.71159&--2.24051&--1.98875 \\ \hline
 1 &   --0.12217 &&&--1.65066& --1.28332& --1.01121&    --2.30409&    --2.03295&    --1.83114\\ \hline
 2&&&& --1.11281&    --0.87216&    --0.66683& --1.90659&--1.7277& --1.57458\\ \hline
 3 &&&& --0.59181& --0.41799&--0.25992& --1.51854& --1.38844 &   --1.26988\\ \hline
 4 &&&&--0.08637&&& --1.13943& --1.03865 &   --0.94383\\ \hline
 5&&&&&&&--0.76877 &--0.68731& --0.60927\\ \hline
 6&&&&&&& --0.40612& --0.33824&--0.27249\\ \hline
\end{tabular}}
\end{table}

\section{Results and discussion}

\begin{figure}[!ht]
\includegraphics[width=0.48\textwidth]{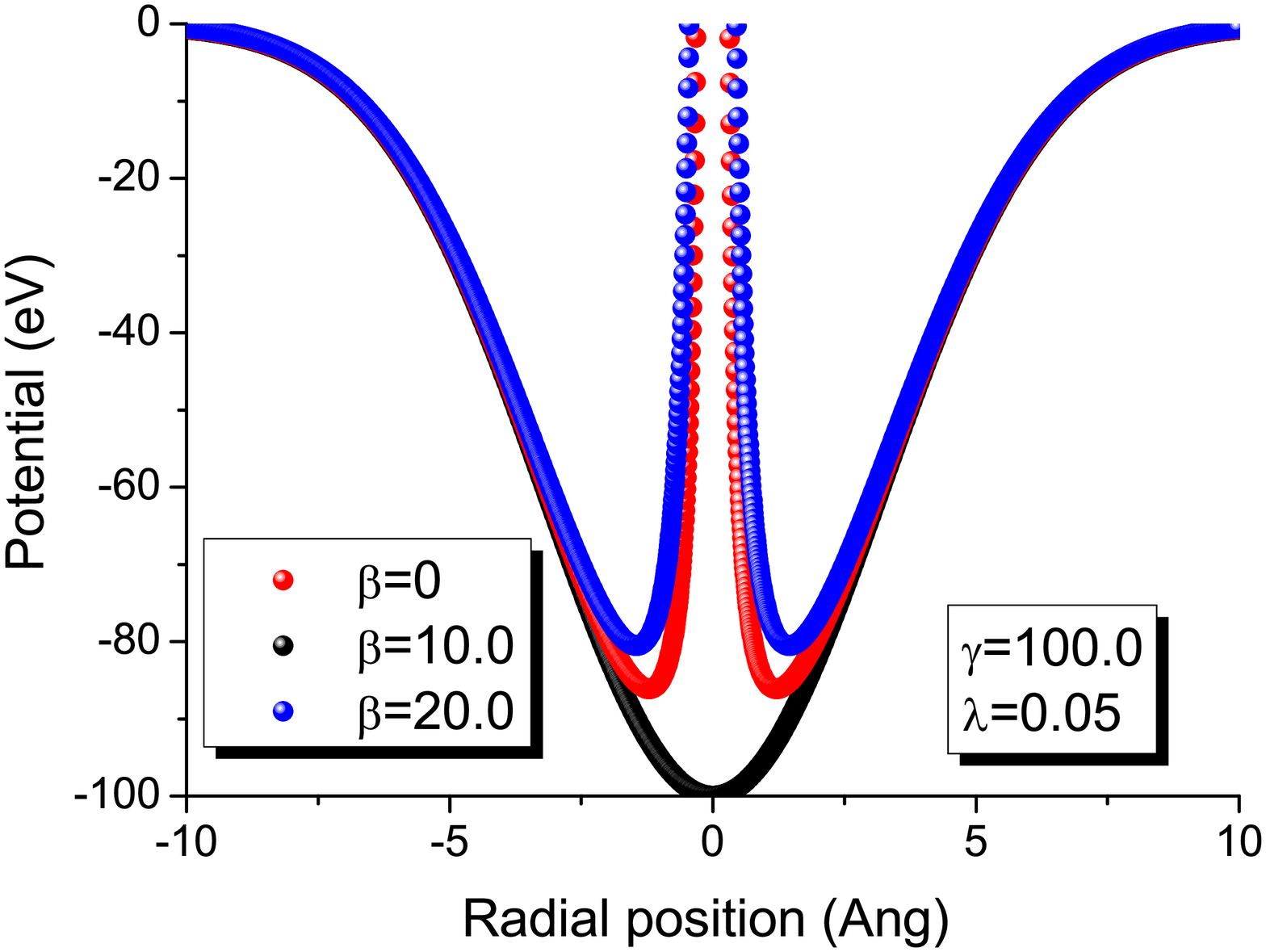}%
\hfill%
\includegraphics[width=0.48\textwidth]{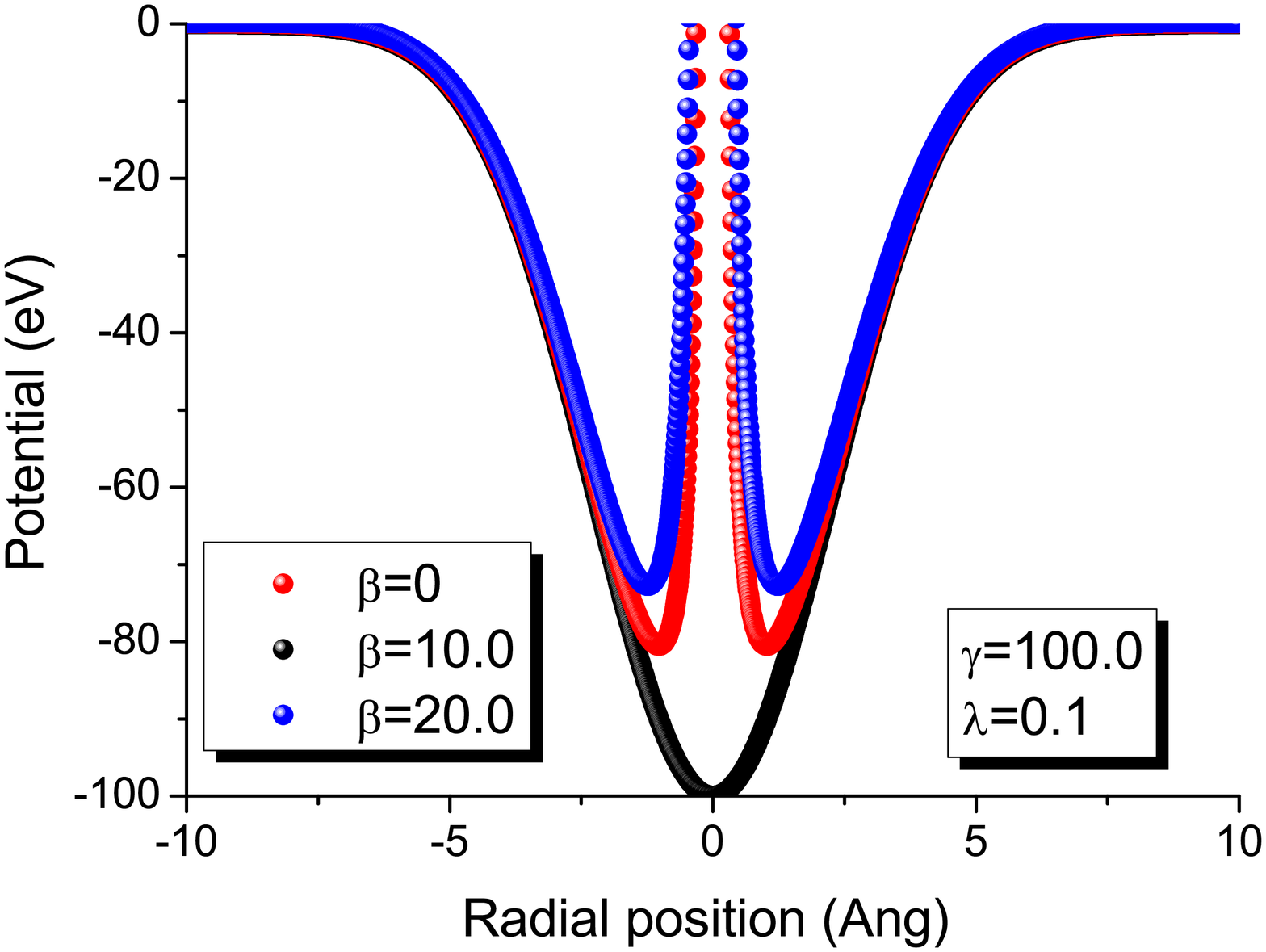}%
\\%
\parbox[t]{0.48\textwidth}{%
\centerline{$\lambda=0.05$}%
}%
\hfill%
\parbox[t]{0.48\textwidth}{%
\centerline{$\lambda=0.1$}
}%
\\
\begin{center}\includegraphics[width=0.48\textwidth]{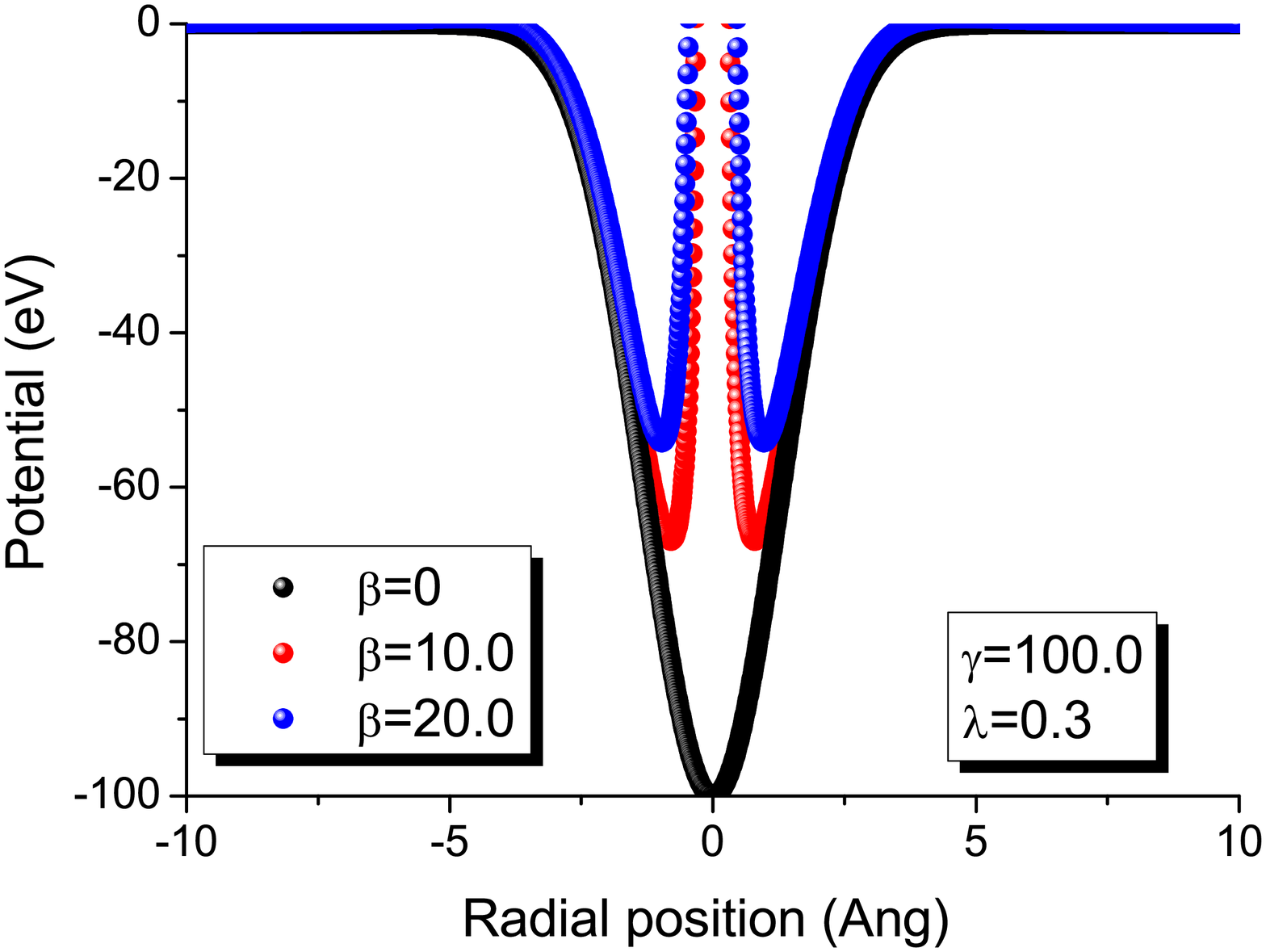}%
\\%
\parbox[t]{0.48\textwidth}{%
\centerline{$\lambda=0.3$}%
}%
\end{center}
\caption{(Color online) The change of the model potential with the change of the
parameters. Each potential can be considered as the one describing hollow
nanoclusters. Here, it should be noted that the units are taken as
eV and Angstrom to provide a possibility to compare the model
parameters with the structure parameters from literature. }
\label{fig:pot}
\end{figure}

The change of the parameters in
the potential of equation~\eqref{gaussian2} refers to the change of the
kind of the molecular structure. As can be seen from figure~\ref{fig:pot}, varying the value of $\lambda$ parameter it is
possible to control the size of the potential well. Furthermore,
using the $\beta$ parameter different from zero, it is possible to
split the gaussian potential into two parts in the positive and the negative
regions. The increase in $\beta$ causes a shrink in the potential.

%
Therefore, the parameters make it possible to produce different model
potentials corresponding to different hollow nanospheres. Tables~\ref{tabl0} and \ref{tabl1} show the eigenvalues for
different parameters. From the tables, it can be concluded that bound
state energy values can be obtained from equation~\eqref{gaussian}.

\section{Conclusions}

We have obtained an analytical form for the energy eigenvalues of
fullerene-type hollow nanospheres by using the technique from
our previous study.  A simple analytical expression (but not an
exact one) is a closed form for the bound states of a modified
attractive Gaussian potential by using supersymmetric perturbation
theory. The analytical expression can be used to see the treatment
of the energy values.
Furthermore, we have shown that this potential can be a good
candidate to model semiconductor quantum dots and to reveal the
electronic and optical properties of the system.

\ukrainianpart

\title{Простий модельний потенціал для порожнистих наносфер}

\author{К. Кьоксал\refaddr{Bitlis}, М. Йонджан\refaddr{Bitlis, Antep}, Б. Гьон'юл\refaddr{Antep}, Б. Гьон'юл\refaddr{Antep}}

\addresses{
\addr{Bitlis} Фізичний факультет, університет Ерена м. Бітліс, 13000, Бітліс, Туреччина
\addr{Antep} Факультет інженерної фізики, університет м. Газіантеп, 27100,
Газіантеп, Туреччина
}

\makeukrtitle

\begin{abstract}
\tolerance=3000
Запропоновано новий модельний потенцiал для опису порожнистих наносфер, таких
як фулерен i молекулярнi структури, та для отримання їхнiх електронних
властивостей. Дано замкнений аналiтичний розв’язок вiдповiдної процедури в
рамках суперсиметричної теорії збурень.

\keywords електронна структура, модельний потенціал, аналітичний
розв'язок

\end{abstract}

\end{document}